\documentclass[10pt,conference]{IEEEtran}
\IEEEoverridecommandlockouts
\usepackage{cite}
\usepackage{amsmath,amssymb,amsfonts}
\usepackage{algorithmic}
\usepackage{graphicx}
\usepackage{textcomp}
\usepackage{xcolor}
\usepackage{hyperref}
\def\BibTeX{{\rm B\kern-.05em{\sc i\kern-.025em b}\kern-.08em
    T\kern-.1667em\lower.7ex\hbox{E}\kern-.125emX}}
\begin{document}

\title{Just Another Quantum Assembly Language (Jaqal)
\thanks{This material was funded by the U.S. Department of Energy, Office
of Science, Office of Advanced Scientific Computing Research Quantum 
Testbed Program.}
}

\author{\IEEEauthorblockN{
B. C. A.  Morrison%
\IEEEauthorrefmark{1}\IEEEauthorrefmark{2},
A. J. Landahl%
\IEEEauthorrefmark{1}\IEEEauthorrefmark{2},
D. S. Lobser%
\IEEEauthorrefmark{1},\\
K. M. Rudinger%
\IEEEauthorrefmark{1},
A. E. Russo%
\IEEEauthorrefmark{1},
J. W. Van Der Wall%
\IEEEauthorrefmark{1},
P. Maunz%
\IEEEauthorrefmark{3}}
\IEEEauthorblockA{
\IEEEauthorrefmark{1}\textit{%
Sandia National Laboratories,
Albuquerque, NM%
} \\
\IEEEauthorrefmark{2}\textit{%
Center for Quantum Information and Control,
University of New Mexico,
Albuquerque, NM
} \\
\IEEEauthorrefmark{3}\textit{%
IonQ, Inc.,
College Park, MD 
} \\
Corresponding author: benmorr@sandia.gov}}

\maketitle

\begin{abstract}
The Quantum Scientific Computing Open User Testbed (QSCOUT) is a trapped-ion
quantum computer testbed realized at Sandia National Laboratories on behalf
of the Department of Energy's Office of Science and its Advanced Scientific
Computing (ASCR) program. Here we describe Jaqal, for Just another quantum
assembly language, the programming language we invented to specify programs
executed on QSCOUT.
Jaqal is useful beyond QSCOUT---it can support mutliple hardware targets
because it offloads gate names and their pulse-sequence definitions to
external files.
We describe the capabilities of the Jaqal language, our approach in
designing it, and the reasons for its creation. To learn more about QSCOUT,
Jaqal, or JaqalPaq, the metaprogramming Python package we developed for
Jaqal, please visit \href{https://qscout.sandia.gov}{qscout.sandia.gov},
\href{https://gitlab.com/jaqal}{gitlab.com/jaqal}, or send an e-mail to
\href{mailto:qscout@sandia.gov}{qscout@sandia.gov}. 
\end{abstract}


\section{Introduction}
QSCOUT is the Quantum Scientific Computing Open User Testbed, a
trapped-ion quantum computer testbed realized at Sandia National
Laboratories on behalf of the Department of Energy's Office of Science
and its Advanced Scientific Computing Research (ASCR) program. As an
open user testbed, QSCOUT provides the following to its users:

\begin{itemize}
\item
  \textbf{Transparency}: Full implementation specifications of the
  underlying native trapped-ion quantum gates.
\item
  \textbf{Extensibility}: Pulse definitions can be programmed to
  generate custom trapped-ion gates.
\item
  \textbf{Schedulability}: Users have full control of sequential and
  parallel execution of quantum gates.
\end{itemize}

In order to provide these features, we need a quantum assembly language
designed around both the flexibility and the detailed control these goals require.
We considered a wide variety of existing languages. However, they either
were lacking on one or more of these points, or were too focused on a
particular hardware model.  We decided to create our own language to meet
our needs.  Due to the proliferation of such languages in this fledgling
field, we named ours \textbf{Just Another Quantum Assembly Language}, or
\textbf{Jaqal}.  We combined key advantages of existing languages to create
one that is flexible enough to target a wide variety of architectures while
supporting the needs of our current hardware.

\subsection{QSCOUT Hardware 1.0}\label{qscout-hardware}

The first version (1.0) of the QSCOUT hardware realizes a single
register of qubits stored in the hyperfine clock states of trapped
${}^{171}$Yb${}^+$ ions arranged in a one-dimensional chain. Single and multi-qubit
gates are realized by tightly focused laser beams that can address
individual ions. The native operations available on this hardware
are the following:

\begin{itemize}
\item
  Preparation and measurement of all qubits in the $z$ basis.
\item
  Parallel single-qubit rotations about any axis in the equatorial plane
  of the Bloch sphere.
\item
  The M{\o}lmer--S{\o}rensen two-qubit gate \cite{b4} between any pair of qubits, in
  parallel with no other gates.
\item
  Single-qubit $Z$ gates executed virtually by adjusting the reference
  clocks of individual qubits.
\end{itemize}

Importantly, QSCOUT 1.0 does not support measurement of a subset of the
qubits.  This limitation is because the resonance fluorescence measurement
process destroys the quantum states of all qubits in the ion chain.
Feedback is therefore also unsupported, since there are no quantum states
onto which feedback can be applied. Future versions of the QSCOUT hardware
will support feedback.

QSCOUT 1.0 uses Jaqal
to specify the quantum programs it executes. On QSCOUT 1.0,
every quantum computation starts with preparation of every qubit in the
$|0\rangle$ state. Then it executes a
sequence of parallel and serial single- and two-qubit gates. After
this, it executes a simultaneous measurement of all qubits in the $z$
basis, returning the result as a binary string. This sequence of
prepare-all/do-gates/measure-all can be repeated multiple times in a
Jaqal program, if desired. However, any adaptive program that uses the
results of one such sequence to issue a subsequent sequence must be done
with metaprogramming, because Jaqal does not currently support feedback.
Once the QSCOUT platform supports classical feedback, Jaqal will be
extended to support it as well.

\subsection{Language Goals}\label{language-goals}

To realize our objectives, the Jaqal quantum assembly language (QASM) fulfills the
following requirements. While many of them were inspired by other
languages' design choices, this particular set was not available
in any mainstream language. Thus, we developed Jaqal to combine the
features most relevant to our specific application, and ideally to a variety
of future platforms with similar goals.

\begin{itemize}
\item
  Jaqal fully specifies the allocation of qubits within the quantum
  register, which \emph{cannot} be altered during execution. This explicit
  specification of which qubits will be used can be found in low-level languages
  like OpenQASM \cite{b3} and Cirq \cite{b2}, but is lacking from many higher-level languages
  which often instead have language constructs for allocating ``clean" qubits in
  a known state or ``dirty" qubits in an unknown one \cite{b8}.
\item
  Jaqal requires the scheduling of serial and parallel gate
  sequencing to be fully and explicitly specified. Many quantum languages
  leave scheduling either completely unspecified, or constrained only by
  ``barrier" statements, which prevent reordering of gates across them but
  allow gates between them to be executed in any order and/or simultaneously.
  Cirq \cite{b2} is a notable exception, allowing circuits to be split into ``moments" in
  which all gates are executed in parallel, with each moment executed
  sequentially after the last. However, Jaqal's scheduling is even more flexible,
  allowing a sequence of gates to be placed in parallel with a single (longer
  duration) gate on a different qubit, for example.

\item
  Jaqal can execute any native (built-in or custom) gate specified in
  any Gate Pulse File it references. There are many standards for
  pulse-level quantum programming, such as OpenPulse \cite{b5}, already available.
  However, these standards do not integrate well with assembly-level 
  languages. You cannot, for example, define a gate at the pulse level in
  OpenPulse and then call that gate from a OpenQASM program alongside
  the built-in native gates. Jaqal allows custom gates to be defined, and their
  implementation encapsulated so their use in a Jaqal program is identical
  to that of a built-in gate.
\item
  Jaqal can be used to define composite gate `macros' which are implemented
  by arbitrary parallel and/or serial combinations of native gates.
  This is a common feature across both QASM-like languages \cite{b3} and other
  quantum programming languages \cite{b7,b8}; Jaqal's syntax is very similar
  to that used by other QASM-like languages.
\item
  Jaqal's built-in execution flow control is sufficient to concisely express common
  benchmarking circuit patterns such as repeated gate set tomography germs \cite{b1}, while also being
  restricted enough to guarantee halting in bounded time.
  Some form of flow control is present in most quantum languages, usually in
the form of 
  conditional execution. Looping constructs like Jaqal's are rarer; Quil does
  have loops, implemented via jump instructions \cite{b7}, but---unlike Jaqal---its jump
  instructions can lead to non-halting programs. Q\# similarly has repeat-until
  and while loops that
  can run for unbounded time \cite{b8}. While this is of course
  necessary for truly universal computation, in practice we are interested
  only in programs of bounded (perhaps polynomial) execution time.
  Jaqal's loops, which run for a fixed number of iterations, guarantee this.
\end{itemize}

While Jaqal is built upon lower-level pulse definitions in Gate Pulse
Files, it is the lowest-level QASM programming language exposed to users in
QSCOUT. As an assembly language, it has been designed with metaprogramming
(specifically generative programming) in mind from the start.  While one can 
write Jaqal directly---and we do not discourage doing
so---we expect most Jaqal code will be machine-generated. This significantly
reduces the burden on the relatively limited classical
computing hardware attached directly to the QSCOUT testbed: all classical
computations are necessarily moved out of the runtime execution of Jaqal and
into a metaprogram that runs on a more sophisticated classical
computer. This means that we omit several features from Jaqal that are
popular in other quantum computing languages, with the expectation that
programs (written in a classical programming language) that generate Jaqal
code can easily replicate those features, at least from the developer's
perspective.  For example, we do not include any form of classical
arithmetic features; if a gate parameter such as a rotation angle needs to
be calculated from known values rather than specified directly, that
calculation occurs at code-generation time rather than at runtime.  To
assist users in this, we have provided a Python package containing tools for writing
metaprograms to generate Jaqal code, called JaqalPaq \cite{b9}.
We anticipate that users will also develop their own
higher-level programming languages that compile down to Jaqal.

\section{Gate Pulse File}\label{gate-pulse-file}

The pulses that implement built-in or custom gates are
defined in a \textbf{\emph{Gate Pulse File (GPF)}}. Eventually, users
will be able to write their own Gate Pulse Files for QSCOUT 1.0, but that capability is not
available in our initial software release \cite{b9}. However, users are
always
free to specify composite gates by defining them as sub-circuit
\hyperref[macro-statement]{macros}. Additionally, user can work 
in collaboration with Sandia scientists to supply their own 
pulse sequences for execution on the QSCOUT hardware. Furthermore, we have
have provided a Gate Pulse File for
the built-in gates of the QSCOUT 1.0 platform, which, other than the
M{\o}lmer-S{\o}rensen gate \cite{b4}, are standard quantum gates as
described in Ref.~\cite{b6}:\footnote{All angles in Jaqal commands are
specified as 64-bit floating point numbers, but are converted by QSCOUT 1.0
hardware to a number between $-2\pi$ and $2\pi$, inclusive, to 40 bits of
precision.}
\begin{itemize}
\item
 \verb|prepare_all| prepares each qubit in the register in the $|0\rangle$
state in the computational ($z$) basis.
\item
 \verb|measure_all| measures each qubit in the register in the computational ($z$) basis.
\item
 \verb|Rx <qubit> <angle>|, \verb|Ry <qubit> <angle>|, and \verb|Rz <qubit> <angle>| rotate the qubit state counterclockwise by an arbitrary angle around the $x$, $y$, or $z$ axis respectively.
\item
 \verb|Px <qubit>|, \verb|Py <qubit>|, and \verb|Pz <qubit>| rotate the qubit state by $\pi$ around the respective axes.
\item
 \verb|Sx <qubit>|, \verb|Sy <qubit>|, and \verb|Sz <qubit>| rotate the qubit state by $\pi/2$ counterclockwise around the respective axes.
\item
 \verb|Sxd <qubit>|, \verb|Syd <qubit>|, and \verb|Szd <qubit>| rotate the qubit state by $\pi/2$ clockwise around the respective axes.
\item
 \verb|MS <qubit> <qubit> <phi> <theta>| is the general two-qubit
M{\o}lmer-S{\o}rensen gate \cite{b4}
 \[\exp\left(-i\left(\frac{\theta}{2}\right)(\cos \varphi X + \sin \varphi Y)^{\otimes 2}\right).\]
\item
 \verb|Sxx <qubit> <qubit>| is the XX-type M{\o}lmer-S{\o}rensen gate with
\(\varphi = 0\) and \(\theta = \pi/2\).
\end{itemize}

We also include idle gates with the same duration as each single- and
 two-qubit gate. While it is not necessary to explicitly insert an idle gate on idling qubits in a
parallel block, these explicit idle gates give the user even more control of the scheduling
of gate execution, and are meant to be used for performance testing and evaluation.

Porting Jaqal to another hardware platform will require the
construction of a GPF that specifies the implementation of its native gates,
whether that is in the form of laser pulses or other control signals.  For
example, a superconducting quantum computer's Gate Pulse File might contain
definitions of a controlled-not gate and a parameterized arbitrary one-qubit
rotation, as well as single-qubit measurements and preparations, in terms of
microwave pulses.  By encapsulating all of the details of the implementation
of gates, state preparation, and measurement instructions into the Gate
Pulse File,
we make Jaqal's core functionality independent of QSCOUT's ion-trap
hardware.

\section{Jaqal Syntax}\label{jaqal-syntax}

A Jaqal file consists of gates and metadata making those gates easier to
read and write. The gates that are run on the machine can be
deterministically computed by inspection of the source text. This
implies that there are no conditional statements at this level. This
section will describe the workings of each statement type.

Whitespace is largely unimportant except as a separator between
statements and their elements. If it is desirable to put two statements
on the same line, a `;' separator may be used. In a parallel block, the
pipe (`\textbar{}') must be used instead of the `;'. Like the semicolon,
however, the pipe is unnecessary to delimit statements on different
lines. Both Windows and Linux newline styles will be accepted.

\subsection{Identifiers}\label{identifiers}

Gate names and qubit names have the same character restrictions. Similar
to most programming languages, they may contain, but not start with,
numerals. They are case sensitive and may contain any non-accented Latin
character plus the underscore. Identifiers cannot be any of the keywords
of the language.

\subsection{Comments}\label{comments}

C/C++ style comments are allowed and treated as whitespace. A comment
starting with `//' runs to the end of the current line, while a comment
with `/*' runs until a `*/' is encountered. These comments do not nest,
which is the same behavior as C/C++.

\subsection{Header Statements}\label{header-statements}

A properly formatted Jaqal file comprises a header and body section. All
header statements must precede all body statements. The order of header
statements is otherwise arbitrary except that all objects must be
defined before their first use.

\subsubsection{Register Statement}\label{register-statement}

A register statement declares how many qubits the user intends to use and
how they will be referred to in the file.  If the machine cannot supply this
number of qubits, then the entire program is rejected immediately.

The following line declares a register named \texttt{q} which holds seven
qubits.

\begin{verbatim}
register q[7]
\end{verbatim}

\subsubsection{Map Statement}\label{map-statement}

While it is sufficient to refer to qubits by their offset in a single
register, it is more convenient to assign names to individual qubits.
The map statement effectively provides an alias to a qubit or array of
qubits under a different name. The following lines declare the single
qubit \texttt{q{[}0{]}} to have the name \texttt{ancilla} and the array
\texttt{qubits} to be an alias for \texttt{q}. Array indices start with
0.

\begin{verbatim}
register q[3]
map ancilla q[0]
map qubits q
\end{verbatim}

The map statement will also support Python-style slicing. In this case,
the map statement always declares an array alias. In the following line,
we relabel every other qubit to be an ancilla qubit, starting with index
1.

\begin{verbatim}
register q[7]
map ancilla q[1:7:2]
\end{verbatim}

After this instruction, \texttt{ancilla{[}0{]}}, \texttt{ancilla{[}1{]}}, and \texttt{ancilla{[}2{]}}
correspond to \texttt{q{[}1{]}}, \texttt{q{[}3{]}}, and \texttt{q{[}5{]}}, respectively.

\subsubsection{Let Statement}\label{let-statement}

We allow identifiers to replace integers or floating point numbers for
convenience. There are no restrictions on capitalization. An integer
defined in this way may be used in any context where an integer literal
is valid and a floating point may similarly be used in any context where
a floating point literal is valid. Note that the values are constant,
once defined.

Example:

\begin{verbatim}
let total_count 4
let rotations 1.5
\end{verbatim}

\subsection{Body Statements}\label{body-statements}

\subsubsection{Gate Statement}\label{gate-statement}

Gates are listed, one per statement, meaning it is terminated either by
a newline or a separator. The first element of the statement is the gate
name followed by the gate's arguments which are whitespace-separated
numbers or qubits. Elements of quantum registers, mapped aliases, and
local variables (see the section on \hyperref[macro-statement]{macros}) may
be freely interchanged as qubit arguments to each gate. The names of the
gates are fixed but determined in the Gate Pulse File, except for
macros. The number of arguments (``arity'') must match the expected
number. The following is an example of what a 2-qubit gate may look
like.

\begin{verbatim}
register q[3]
map ancilla q[1]
Sxx q[0] ancilla
\end{verbatim}

The invocation of a macro is treated as completely equivalent to a gate
statement.

\subsubsection{Gate Block}\label{gate-block}

Multiple gates and/or macro invocations may be combined into a single
block. This is similar, but not completely identical, to how C or
related languages handle statement blocks. Macro definitions and header
statements are not allowed in gate blocks. Additionally, statements such
as macro definitions or loops expect a gate block syntactically and are
not satisfied with a single gate, unlike C.

Two different gate blocks exist: sequential and parallel. Sequential
gate blocks use the standard C-style `\{\}' brackets while parallel
blocks use angled `\textless{}\textgreater{}' brackets, similar to C++
templates. This choice was made to not conflict with `{[}{]}' brackets,
which are used in arrays, and to reserve `()' for possible future use.
In a sequential block, each statement, macro, or gate block waits for
the previous to finish before executing. In a parallel gate block, all
operations are executed at the same time. It is an error to request
parallel operations that the machine is incapable of performing, however
it is not syntactically possible to forbid these as they are determined
by hardware constraints which may change with time.

The Jaqal language does not have a barrier statement, as many other
quantum assembly languages do, that specifies to the execution environment
which gates should not be re-ordered to be executed simultaneously.
Jaqal gates will be executed simultaneously if and only if the user places them
in a parallel block with each other, with no re-ordering at runtime, so no
such statement is necessary.

\hyperref[loop-statement]{Looping statements} are allowed inside
sequential blocks, but not inside parallel blocks. Blocks may be
arbitrarily nested so long as the hardware can support the resulting
sequence of operations. Blocks may not be nested directly within other
blocks of the same type.

The following statement declares a parallel block with two gates.

\begin{verbatim}
< Sx q[0] | Sy q[1] >
\end{verbatim}

This does the same but on different lines.

\begin{verbatim}
<
    Sx q[0]
    Sy q[1]
>
\end{verbatim}

Here is a parallel block nested inside a sequential one.

\begin{verbatim}
{ Sxx q[0] q[1]; < Sx q[0] | Sy q[1] >; }
\end{verbatim}

And sequential blocks may be nested inside parallel blocks.

\begin{verbatim}
< Px q[0] | { Sx q[1] ; Sy q[1] } >
\end{verbatim}

\subsubsection{Timing within a parallel
block}\label{timing-within-a-parallel-block}

If two gates are in a parallel block but have different durations
(\emph{e.g.}, two single-qubit gates of different length), the default
behavior is to \emph{start} each gate within the parallel block
simultaneously. The shorter gate(s) will then be padded with idles until
the end of the gate block. For example, the command

\begin{verbatim}
< Rx q[1] 0.1 | Sx q[2] >
\end{verbatim}

results in the \texttt{Rx} gate on \texttt{q{[}1{]}} with angle 0.1
radians and \texttt{Sx} gate on \texttt{q{[}2{]}} both starting at the
same time; the \texttt{Rx} gate will finish first and \texttt{q{[}1{]}}
will idle while the \texttt{Sx} gate finishes. Gate Pulse Files will allow users to 
define their own scheduling within parallel blocks
(\emph{e.g.}, so that gates all \emph{finish} at the same time instead).
When using the QSCOUT GPF, explicit idle gates with durations matching each
native gate are also available for manual padding.

\hyperdef{}{macro-statement}{\subsubsection{Macro
Statement}\label{macro-statement}}

A macro can be used to treat a sequence of gates as a single gate. Gates
inside a macro can access the same qubit registers and mapped aliases at
the global level as all other gates, and additionally have zero or more
arguments which are visible. Arguments allow the same macro to be
applied on different combinations of physical qubits, much like a
function in a classical programming language.

A macro may use other macros that have already been declared. A macro
declaration is complete at the \emph{end} of its code block. This implies
that recursion is impossible. It also implies that macros can only reference
other macros created earlier in the file. Because Jaqal has no conditional
statements, if this restriction were not in place, recursion would create
an infinite loop.

A macro is declared using the \texttt{macro} keyword, followed by the
name of the macro, zero or more arguments, and a code block. Unlike C, a
macro must use a code block, even if it only has a single statement.

The following example declares a macro.

\begin{verbatim}
macro foo a b {
    Sx a
    Sxx a q[0]
    Sxx b q[0]
}
\end{verbatim}

To simplify parsing, a line break is not allowed before the initial
`\{', unlike C. However, statements may be placed on the same line
following the `\{'.

\hyperdef{}{loop-statement}{\subsubsection{Loop
Statement}\label{loop-statement}}

A gate block may be executed for a fixed number of repetitions using the
loop statement. The loop statement is intentionally restricted to
running for a fixed number of iterations. This ensures it is easy to
deterministically evaluate the runtime of a program. Consequently, it is
impossible to write a program which will not terminate.

The following loop executes a sequence of statements seven times.

\begin{verbatim}
loop 7 {
    Sx q[0]
    Sz q[1]
    Sxx q[0] q[1]
}
\end{verbatim}

The same rules apply as in macro definitions: `\{' must appear on the
same line as \texttt{loop}, but other statements may follow on the same
line.

Loops may appear in sequential gate blocks, but not in parallel gate
blocks.

\section{Extensibility}\label{extensibility}

As Jaqal, and the QSCOUT project more broadly, have extensibility as
stated goals, it is important to clarify what is meant by this term.
Primarily, Jaqal offers extensibility in the gates that can be
performed. This will occur through the Gate Pulse File and the use of
macros to define composite gates that can be used in all contexts permitted
for native gates. Jaqal will be incrementally improved as new hardware
capabilities come online and real-world use identifies areas for
enhancement. The language itself, however, is not intended to have many
forms of user-created extensibility as a software developer might
envision the term. Features we do not intend to support include, but are
not limited to, pragma statements, user-defined syntax, and a foreign
function interface (\textit{i.e.,~}using custom C or Verilog code in a Jaqal
file).

\hyperdef{}{data-output-format}{\section{Data Output
Format}\label{data-output-format}}

When successfully executed, a single Jaqal file will generate a single
ASCII text file (Linux line endings) in the following way:

\begin{enumerate}
\def\labelenumi{\arabic{enumi}.}
\item
  Each call of \texttt{measure\_all} at runtime will add a new line of
  data to the output file. (If \texttt{measure\_all} occurs within a
  \texttt{loop} (or nested loops), then multiple lines of data will be
  written to the output file, one for each call of \texttt{measure\_all}
  during execution.)
\item
  Each line of data written to file will be a single bitstring, equal in
  length to the positive integer passed to \texttt{register} at the
  start of the program.
\item
  Each bitstring will be written in least-significant bit order (little
  endian).
\end{enumerate}

For example, consider the program:

\begin{verbatim}
register q[2]

loop 2 {
    prepare_all
    Px q[0]
    measure_all
}

loop 2 {
    prepare_all
    Px q[1]
    measure_all
}
\end{verbatim}

Assuming perfect execution, the output file would read as:

\begin{verbatim}
10
10
01
01
\end{verbatim}

While this output format will be ``human-readable'', it may nevertheless be
unwieldy to work with directly. To help with this, we included a
Python-based parser in the JaqalPaq metaprogramming package to aid users in
manipulating output data \cite{b9}.

\section{Metaprogramming}\label{possible-future-capabilities}

\subsection{Compile-Time Classical
Computation}\label{compile-time-classical-computation}

Performing classical computations at compile-time, before the program is
sent to the quantum computer, can vastly increase the expressiveness of
the language.  We provide three examples of this here.

\subsubsection{Calculating Gate Angles}
Jaqal does not have built-in arithmetic functions, meaning that gate
parameters cannot be calculated by a Jaqal program itself.
For example, consider the following program,
\emph{which is not currently legal Jaqal code:}

\begin{verbatim}
register q[1]

let pi 3.1415926536

loop 100 {
    prepare_all; Ry q[0] pi/32; measure_all
    prepare_all; Ry q[0] pi/16; measure_all
    prepare_all; Ry q[0] pi/8; measure_all
}
\end{verbatim}

If writing such a Jaqal program by hand, a user can define constants as needed
to store the computed values:

\begin{verbatim}
register q[1]

let pi_32   0.09817477042
let pi_16   0.1963495408
let pi_8    0.3926990817

loop 100 {
    prepare_all; Ry q[0] pi_32; measure_all
    prepare_all; Ry q[0] pi_16; measure_all
    prepare_all; Ry q[0] pi_8; measure_all
}
\end{verbatim}

However, if a user generates a Jaqal program via a high-level language, then
he or she can include the calculations inline and have the metaprogram
automatically substitute the results, as in the following pseudocode block:

\pagebreak

\begin{verbatim}
q = register("q", 1)
loop(100, gate("prepare_all"), 
     gate("Ry", q[0], pi/32), 
     gate("measure_all"), 
     gate(), ...)
\end{verbatim}

Assuming suitable definitions of the register, loop, and gate calls in the 
metalanguage environment, running that metaprogram would then generate a 
Jaqal program:

\begin{verbatim}
register q[1]

loop 100 {
    prepare_all;
    Ry q[0] 0.09817477042; 
    measure_all;
    ...
}
\end{verbatim}

\subsubsection{Macro Definition}
Another example of a case where compile-time classical computation could
be useful is in macro definitions. For example, if a user wished to define a
macro for a controlled-$z$ rotation in terms of a (previously-defined)
CNOT macro, the following \emph{would not be legal Jaqal code:}

\begin{verbatim}
...
macro CNOT control target { ... }

macro CRz control target angle {
    Rz target angle/2
    CNOT control target
    Rz target -angle/2
    CNOT control target
}
...
CRz q[0] q[1] 0.7853981634;
...
\end{verbatim}

This is because both \texttt{angle/2} and \texttt{-angle/2} are classical
computations that are not permitted in Jaqal.  Instead of defining a Jaqal
macro, users could write the relevant gate sequence manually:

\begin{verbatim}
...
Rz q[1] 0.3926990817;
CNOT q[0] q[1];
Rz q[1] -0.3926990817;
RNOT q[0] q[1];
...
\end{verbatim}

However, this makes for unexpressive code. Instead, a user could define the 
controlled Z-rotation using a metalanguage, similarly to the following 
pseudocode:

\begin{verbatim}
procedure CRz (ctrl, target, angle) {
  gate("Rz", target, angle/2)
  gate("CNOT", ctrl, target)
  gate("Rz", target, -angle/2)
  gate("CNOT", ctrl, target)
}
...
CRz(q[0], q[1], 0.7853981634)
...
\end{verbatim}

\subsubsection{Randomized Algorithms}
Another relevant use of classical computation is to generate random numbers
to determine the sequence of gates. Applications of randomized quantum
programs include hardware benchmarking, error mitigation, and some
quantum simulation algorithms. This, too, can be done in this generative
programming paradigm, pre-generating all the random values and
automatically producing Jaqal code to execute the random circuit selected.

\subsection{Run-Time Classical
Computation}\label{run-time-classical-computation}

Users may also wish to execute runtime classical computations based on
intermediate measurement results. For example, in hybrid variational algorithms, a
classical optimizer may use measurement results from one circuit to choose
rotation angles used in the next circuit. In error-correction experiments, a
decoder may need to compute which gates are necessary to restore a state
based on the results of stabilizer measurements. Adaptive tomography
protocols may need to perform statistical analyses on measurement results to
determine which measurements will give the most information.

Run-time classical computations fall into two main categories: determining
which circuits to run based on measurement results from former circuits, and
determining the gate sequence of a circuit based on intermediate
measurements within a circuit.  Jaqal supports the former, but not the latter,
because QSCOUT 1.0 does not support classical feedback.  We anticipate that
once the QSCOUT hardware does support classical feedback, we will extend
Jaqal to support it as well.

However, use cases like adaptive tomography and variational algorithms are
possible with the current hardware, and can be implemented via
metaprogramming techniques. After running a Jaqal program  on the QSCOUT
hardware, a metaprogram can parse the
\hyperref[data-output-format]{measurement results}, then use that
information to generate a new Jaqal program  to run. This allows for
adaptive and hybrid algorithms to be run without having to execute a numerical
optimization routine or similar such tool on the classical control circuitry
of QSCOUT.
The JaqalPaq package has example variational quantum eigensolver algorithms
for several molecules to demonstrate how this can be done \cite{b9}.

\section*{Acknowledgment}\label{acknowledgements}

Sandia National Laboratories is a multi-mission laboratory managed and
operated by National Technology and Engineering Solutions of Sandia,
LLC, a wholly owned subsidiary of Honeywell International, Inc., for
DOE's National Nuclear Security Administration under contract
DE-NA0003525.


\begin{thebibliography}{00}
\bibitem{b1} R. Blume-Kohout, \textit{et al.}, ``Demonstration of qubit operations below a rigorous fault tolerance threshold with gate set tomography,'' Nature Comm. \textbf{8} (1), 1, (2017).
\bibitem{b2} Cirq: A Python framework for creating, editing, and invoking Noisy Intermediate Scale Quantum (NISQ) circuits, https://github.com/Cirq.
\bibitem{b3} A. W. Cross, L. S. Bishop, J. A. Smolin, and J. M. Gambetta, ``Open quantum assembly language,'' 2017, arXiv:1707.03429, unpublished.
\bibitem{b9} JaqalPaq (Python Jaqal Programming Package), https://gitlab.com/jaqal/jaqalpaq.
\bibitem{b4} K. M{\o}lmer and A. S{\o}rensen, ``Multiparticle entanglement of hot trapped ions." Phys. Rev. Lett. \textbf{82} (9), 1835, (1999).
\bibitem{b5} D. McKay et. al, ``Qiskit backend specifications for OpenQASM and OpenPulse experiments," 2018, arXiv:1809.03452, unpublished.
\bibitem{b6} M. A. Nielsen and I. L. Chuang, Quantum information and quantum computation. Cambridge: Cambridge University Press, 2000.
\bibitem{b7} R. S. Smith, M. J. Curtis, and W. J. Zeng, ``A practical quantum instruction set architecture,'' 2016, arXiv:1608.03355, unpublished.
\bibitem{b8} The Q\# programming language, https://docs.microsoft.com/en-us/quantum/language/?view=qsharp-preview.
\end{thebibliography}
\end{document}